%% file: silaf_proc_LHCb.tex
\documentclass[3p,times,twocolumn]{elsarticle}

\usepackage{ecrc}

\volume{00}

\firstpage{1}

\journalname{Nuclear Physics B Proceedings Supplement}

\runauth{A. Hicheur}

\jid{nuphbp}

\jnltitlelogo{Nuclear Physics B Proceedings Supplement}

\usepackage{amsmath} 
\usepackage{amssymb}
\usepackage{amsfonts}
\usepackage{upgreek} 

\usepackage{xspace}
\usepackage{ifthen} 
\newboolean{articletitles}
\setboolean{articletitles}{true} 

\newboolean{uprightparticles}
\setboolean{uprightparticles}{false} 

\newboolean{inbibliography}
\setboolean{inbibliography}{false} 

\usepackage{mciteplus}

\input{lhcb-symbols-def}

\begin{document}

\begin{frontmatter}



\title{Selection of LHCb results from Run I}


\author{A. Hicheur \\
{\it hicheur@if.ufrj.br }\\
 {\it On behalf of the LHCb collaboration}}

\address{Instituto de F\'isica, Universidade Federal do Rio de Janeiro \\
CP 68528   CEP 21945-970 Rio de Janeiro, RJ, Brasil. \\}

\begin{abstract}
At the eve of the second LHC data taking run, some of the most recent results obtained by the \lhcb collaboration with Run I data are reviewed. Improved measurements on \CP violation, unitary triangle and mixing parameters are shown. Recent progress on physics in the forward region is illustrated by examples picked up in the electroweak physics and beyond Standard Model searches.
\end{abstract}


\end{frontmatter}


\section{Introduction}
\label{}
The \lhcb research programme is principally focused on using heavy flavour decays to probe the intervention of New Physics through the involvement of heavy new particles in loop decays. For this purpose, neutral $B^0_q$ ($q=d,s$) meson oscillations, mediated by box diagrams, and rare decays, proceeding through loop flavour changing neutral currents, come as natural grounds to seek for any deviation from Standard Model-based predictions.~Spectroscopy of heavy-flavoured hadrons is also considered along these research lines.

Beside this, the low transverse momentum threshold, the excellent vertex reconstruction, and the peculiar forward geometry of the \lhcb detector allows for complementary studies to what is performed with \atlas or \cms detectors, on areas such as QCD and Electroweak physics, as well as searches for displaced vertices of heavy decaying particles.

Most of the analyses are based on the full Run I (years 2011 and 2012) data sample, representing 1.0 and 2.0 \invfb of integrated luminosity at $7$ and $8$ TeV center-of-mass energies in $pp$ collisions, respectively. Some of the studies exploit a subsample of this set and are under active preparation of updated results.

\section{\lhcb detector and software}
\begin{figure}[t]
\begin{center}
\includegraphics[width=0.5\textwidth]{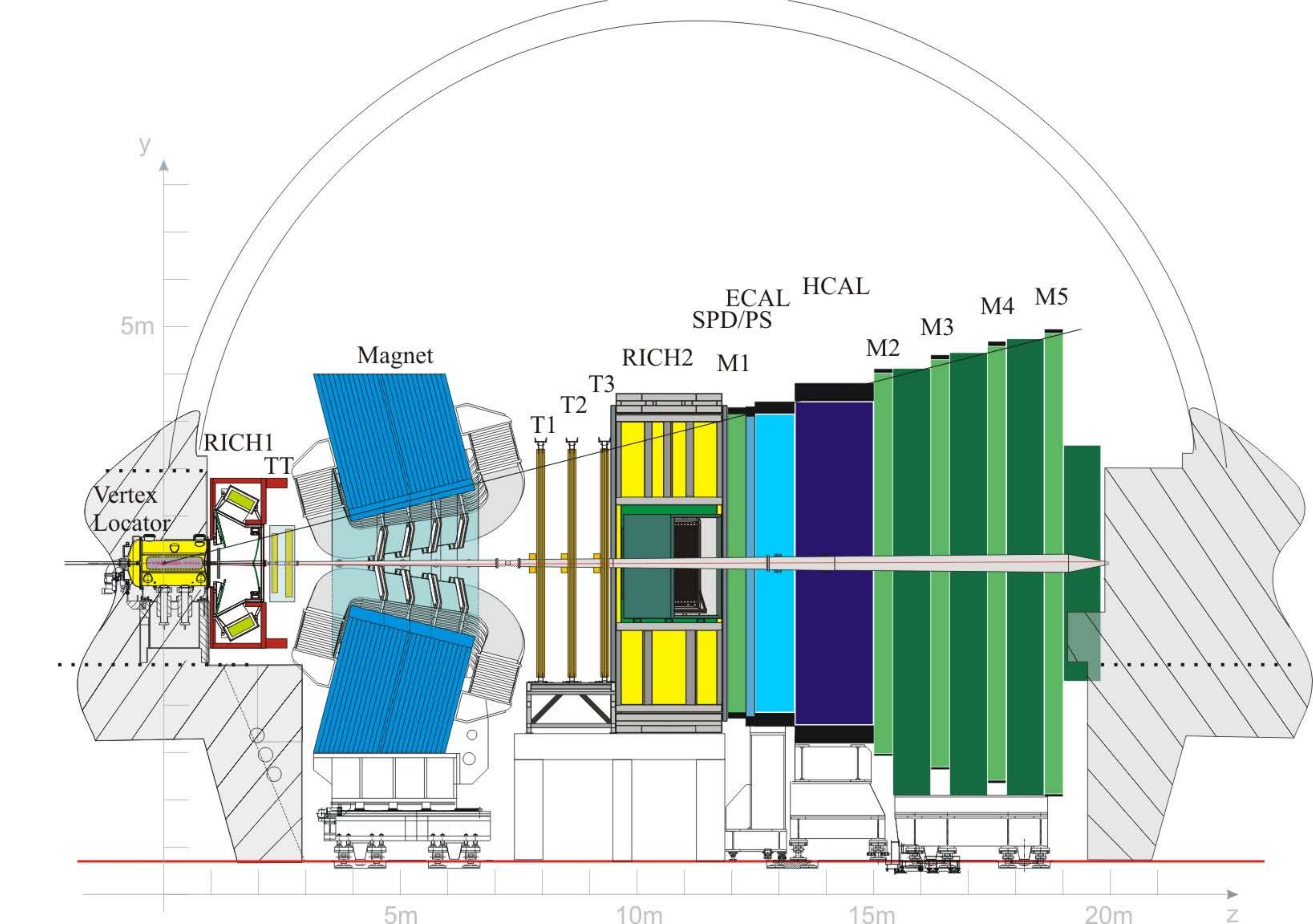}
\end{center}
\caption{Global view of the \lhcb detector.}
\label{Fig:lhcb_layout}
\end{figure}

The \lhcb detector~\cite{Alves:2008td}, Fig.\ref{Fig:lhcb_layout}, is a single-arm forward spectrometer covering the \mbox{pseudorapidity} range $2<\eta <5$, specially designed for the study of particles containing $b$ or $c$ quarks.~The detector includes a high precision tracking system, consisting of a silicon-strip vertex detector surrounding the $pp$ interaction region, a large-area silicon-strip detector located upstream of a dipole magnet with a bending power of about $4{\rm\,Tm}$, and three stations of silicon-strip detectors and straw drift tubes placed downstream.~The combined tracking system has momentum resolution $\Delta p/p$ that varies from 0.4\% at 5\gevc to 0.6\% at 100\gevc, and impact parameter~(IP) resolution of 20\mum for tracks with high transverse momentum.~Charged hadrons are identified using two ring-imaging Cherenkov detectors (RICH) \cite{RichPerf}. Photon, electron and hadron candidates are identified by a calorimeter system consisting of scintillating-pad and preshower detectors, an electromagnetic calorimeter and a hadronic calorimeter.~Muons are identified by a system composed of alternating layers of iron and multiwire proportional chambers. 

The trigger~\cite{Aaij:2012me} consists of a hardware stage, based on information from the calorimeter and muon systems, followed by a software stage which applies a full event reconstruction.~Events triggered both on objects independent of the signal, and associated with the signal, are used. In the latter case, the transverse energy of the hadronic cluster is required to be at least 3.5\gev. The software trigger includes more sophisticated requirements and operations, in particular a multivariate algorithm to identify secondary vertices~\cite{Gligorov:2012qt}.

The analyses use simulated events generated by \pythia~8.1~\cite{Sjostrand:2008} with a specific \lhcb configuration~\cite{LHCb-PROC-2010-056}.  Decays of hadronic particles are described by \evtgen~\cite{Lange:2001uf} in which final state radiation is generated using \photos~\cite{Golonka:2005pn}. The interaction of the generated particles with the detector and its response are implemented using the \geant toolkit~\cite{Allison:2006ve,*Agostinelli:2002hh} as described in Ref.~\cite{LHCb-PROC-2011-006}.

\section{$\gamma$ angle with tree and loop decays}
\label{sec:gamma_angle}
The angle $\gamma=\mathrm{arg}\left(\frac{-V_{ub}V_{ud}^*}{V_{cb}V_{cd}^*}\right)$ is the less known angle of the Cabibbo-Kobayashi-Maskawa (CKM) \cite{CKM1,CKM2} unitarity triangle formed by the condition $V_{ub}V_{ud}^*+V_{cb}V_{cd}^*+V_{tb}V_{td}^*=0$. To extract it, tree-level $B^+$-meson decays to a neutral charmed meson $D^0$ and a charged kaon are used, but other variants include $B^0_d$ or $B^0_s$ decays involving $D^0$ or $D_s^+$.

Interferences between allowed and suppressed amplitudes of $B\to DK$ and $D\to f_D$ decays accessed by both $D$ and $\bar{D}$, leading to a unique $f_DK$ final state, are exploited. $f_D$ may be a \CP eigenstate such as $\Kp\Km$ \cite{GLW}, a flavour-specific state such as $\Kp\pim$ \cite{ADS}, a multibody state such as $\KS\pip\pim$ \cite{GGSZ,GLS} or $\Kp\pim\pip\pim$. In all these studies, allowed $B$ and $D$ decays amplitudes are labelled as $A_B$ and $A_D$, while the suppressed are parametrized using amplitude ratios $r_i$, strong phases $\delta_i$ and the $\gamma$ angle, $r_BA_B e^{i(\delta_B-\gamma)}$ and $r_DA_D e^{i\delta_D}$. The quantities $r_D$ and $\delta_D$ are generally obtained from independent studies, while $r_B$, $\delta_B$ and $\gamma$ are extracted through observables such as charge asymmetries, or ratios of $D\to f_D$ allowed and suppressed chains.

\begin{figure}[t]
\begin{center}
\includegraphics[width=0.5\textwidth]{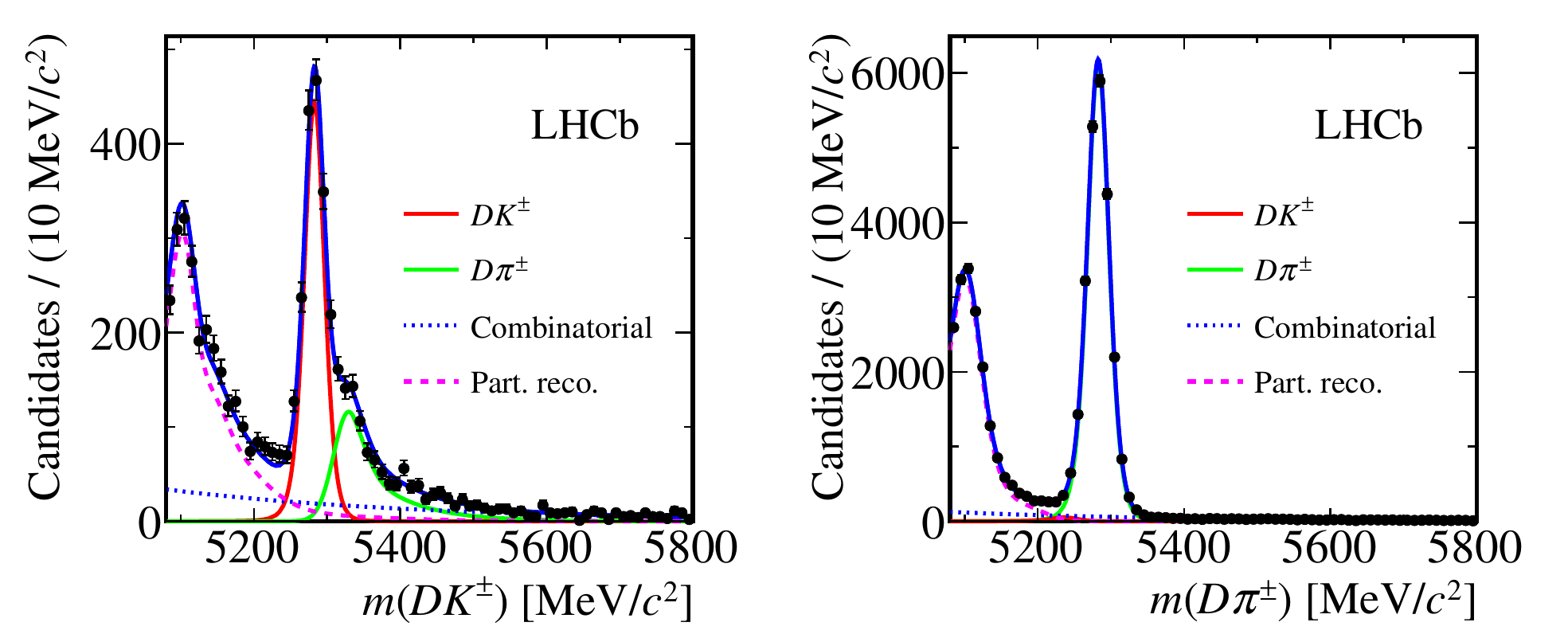}
\includegraphics[width=0.5\textwidth]{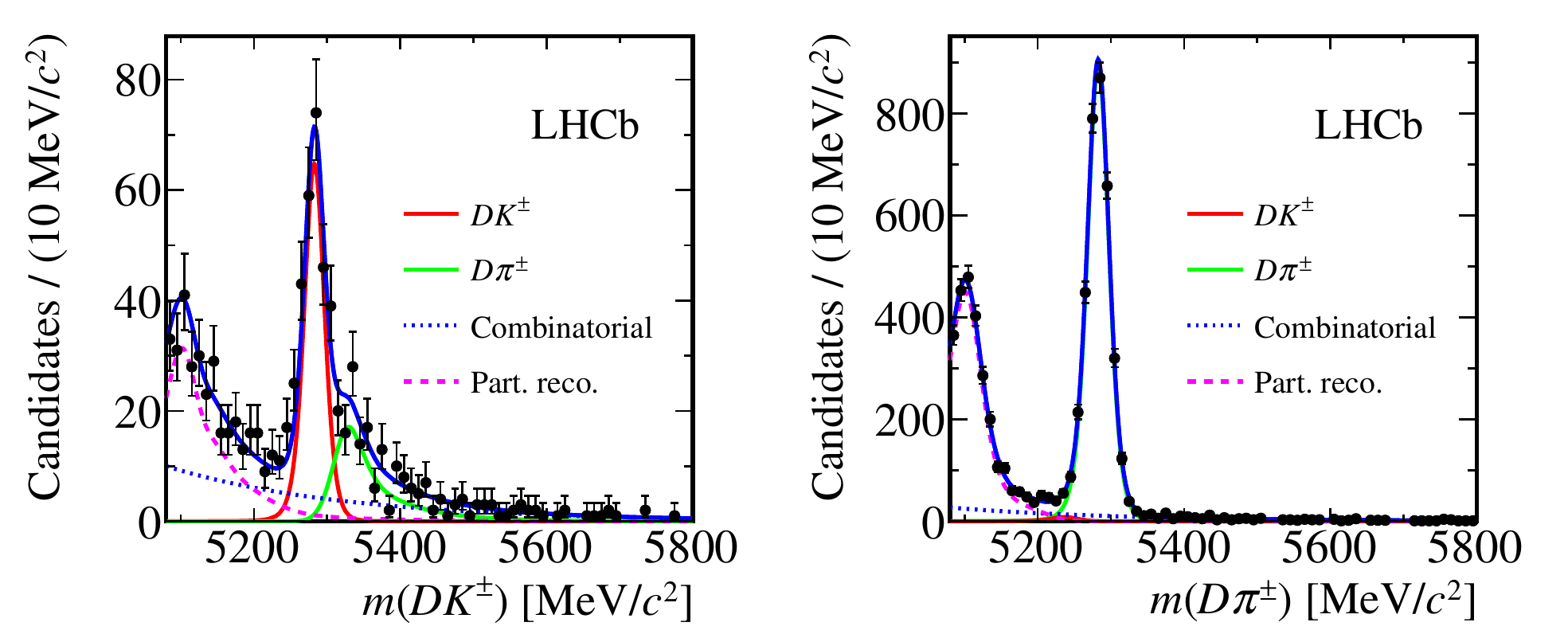}
\end{center}
\caption{Invariant mass distributions of $D^0\Kpm$ (left) and $D^0\pipm$ (right) for $\Bpm\to D^0(\KS \pi^+\pi^-)\Kpm$ (top) and $\Bpm\to D^0(\KS K^+K^-)\Kpm$ (bottom) decay chains.}
\label{Fig:kshhk_DD}
\end{figure}

The \lhcb collaboration has investigated many of these decay chains, among which the recent results on the charged modes $\Bpm\to D^0(\KS h^+h^-)\Kpm$ ($h=\pi$ or $K$) \cite{lhcb_bdk_kshh}, $\Bpm\to D^0(\KS\Kp\pim)\Kpm$  \cite{lhcb_bdk_kskpi}, and neutral modes $\B^0\to D^0(\Km\pip)K^{*0}$ \cite{lhcb_b0dkstar} and $\Bs\to D_s^+ K^-$ \cite{lhcb_bsdsk}, this latter representing the first time a \Bs decay of this type is used to extract $\gamma$. 
In the case of $\Bpm\to D^0(\KS h^+h^-)\Kpm$, which fits to the signal are represented in Fig.\ref{Fig:kshhk_DD}, model-independent distributions of $D^0\to\KS h^+h^-$ decays are used to describe $r_D$ and $\delta_D$ \cite{cleoc_dkshh}, and the charge asymmetry distribution permits to infer $\gamma=(62^{+15}_{-14})^\circ$, $r_B=0.080^{+0.019}_{-0.021}$ and $\delta_B=(134^{+14}_{-15})^\circ$. For $\Bs\to D_s^+ K^-$, $\gamma=(115^{+28}_{-43})^\circ$ is obtained after including the most recent measurement of the \Bs mixing phase $-2\beta_s$ in the fit. A combination of all available \lhcb tree-level results yields $\gamma=(72.9^{+9.2}_{-9.9})^\circ$ (frequentist) or $\gamma=(71.9^{+9.9}_{-10.0})^\circ$ (bayesian) \cite{LHCb-CONF-2014-004}.

At loop level, the charged and neutral two-body decays $\Bpm\to\pipm\pi^0$, $\B^0\to\pip\pim$, $\B^0\to\pi^0\pi^0$ and $\Bs\to\Kp\Km$ are used to obtain a measurement of $\gamma$ \cite{LHCb-PAPER-2014-045}.~The observables used in the fit are the branching fractions and the \CP violation parameters of the decays. Since both tree and loop diagrams contribute to the amplitudes, the number of unknowns exceeds the number of observables and assumptions are thus needed. Both isopin symmetry \cite{GL_isospin}, relating $\Bpm\to\pipm\pi^0$ and $\B^0\to\pi^0\pi^0$, and U-spin symmetry \cite{Fleischer}, relating $\B^0\to\pip\pim$ and $\Bs\to\Kp\Km$, are used to constrain the number of unknowns. Non-factorizable and U-spin breaking effects are considered as perturbing factors to amplitude ratios.~Using the measured value of $-2\beta_s$, and allowing for up to 50\% of U-spin breaking, the fit yields $\gamma=(63.5^{+7.2}_{-6.5})^\circ$, which is compatible with the tree-level average mentioned in the previous paragraph.

In the future, as precision increases with more data, any significant mismatch between the two methods of extraction would point to electroweak or TeV-scale effects occuring in loop diagrams.

\section{\CP asymmetries in oscillations and decays}
\begin{figure*}[t]
\begin{center}
\includegraphics[width=0.4\textwidth]{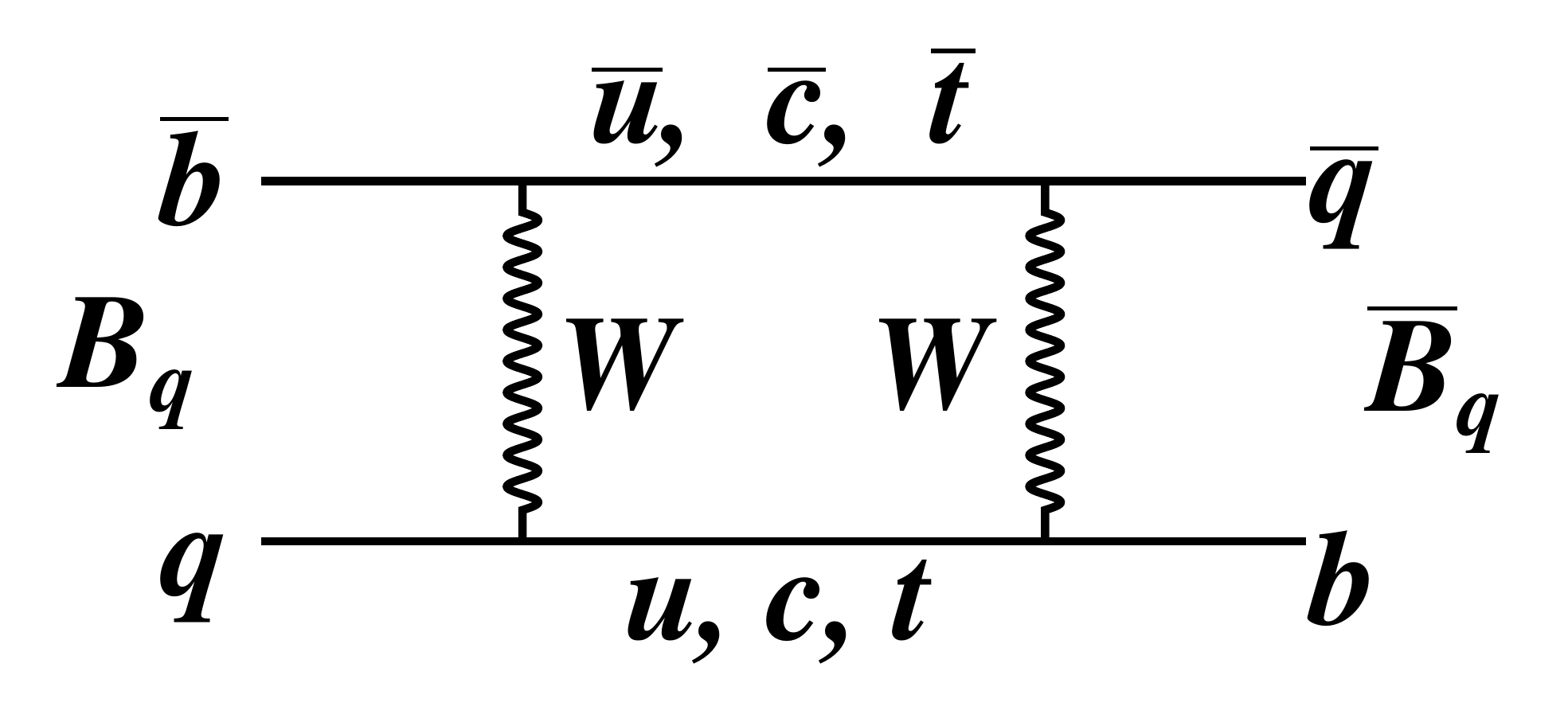}
\includegraphics[width=0.4\textwidth]{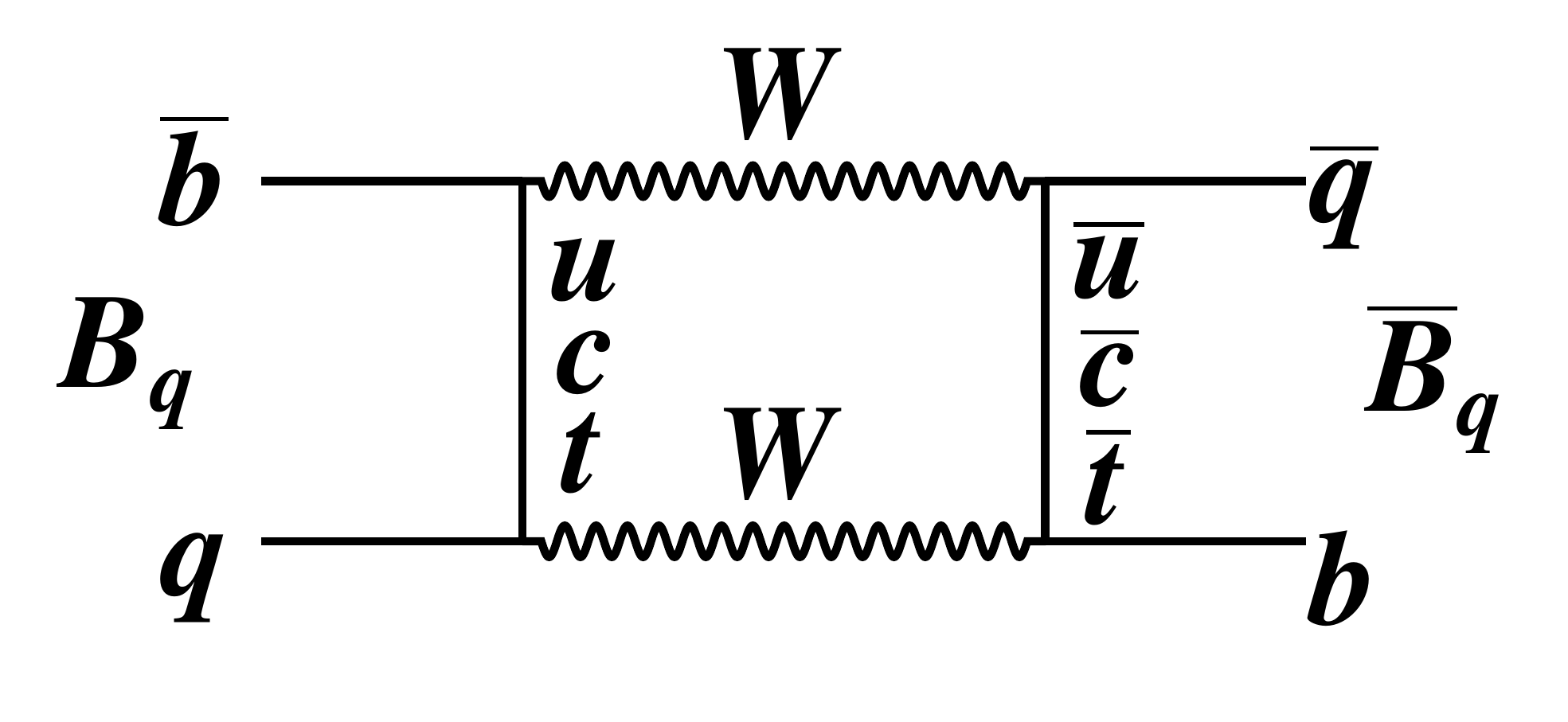}
\end{center}
\caption{Oscillation diagrams of $B_q$ ($q=d,s$) mesons.}
\label{Fig:boxdiag}
\end{figure*}

Oscillations of neutral $B$-mesons are represented by a two-state system governed at short ranges by the diagrams shown in Fig.\ref{Fig:boxdiag}. 
The oscillation dynamics is written as:
\begin{equation}
i\frac{d}{dt}\begin{pmatrix}|B^0_q(t)\rangle \vspace{1mm}\\|\overline B^0_q(t)\rangle\end{pmatrix}=\Big[\begin{pmatrix}M_{11}^q & M_{12}^q \vspace{2mm}\\ M_{12}^{q*} & M_{22}^q\end{pmatrix}-\frac{i}{2}\begin{pmatrix}\Gamma_{11}^q & \Gamma_{12}^q \vspace{2mm}\\\Gamma_{12}^{q*} & \Gamma_{22}^q\end{pmatrix}\Big]\begin{pmatrix}|B^0_q(t)\rangle \vspace{1mm}\\|\overline B^0_q(t)\rangle\end{pmatrix},
\end{equation}
where $M_{ij}^q$ and $\Gamma_{ij}^q$ are elements of the mass and decay matrices. We note $\phi^q_{12}=\mathrm{arg}(-M_{12}^q/\Gamma_{12}^q)$ and $\phi^q_{M}=\mathrm{arg}(M_{12}^q)$.
In the standard model, \CP violation in mixing $\phi^q_{12}$ is expected to be very small and thus an intervention of new heavy particles could be detected through a substantial deviation from the null prediction.\\
For both $\Bd$ and $\Bs$, a way to do it is to use semileptonic (and thus flavour-specific) decays to obtain
\begin{eqnarray}
\nonumber A_{sl}^q &=& \frac{\Gamma(\overline B^0_q \to B^0_q \to f)-\Gamma(B^0_q \to \overline B^0_q \to \overline f)}{\Gamma(\overline B^0_q \to B^0_q \to f)+\Gamma(B^0_q \to \overline B^0_q \to \overline f)}\\
&=& \frac{\Delta\Gamma_q}{\Delta m_q}\tan(\phi^q_{12})
\end{eqnarray}
where $\Delta\Gamma_q$ and $\Delta m_q$ are the width and mass differences between the mass eigenstates. In practice, \lhcb analyses measure the time-dependent yield asymmetry of the decay states \cite{LHCb-PAPER-2013-033,LHCb-PAPER-2014-053}:
\begin{eqnarray}
\frac{N(B^0_q \to f,t)-N(\overline B^0_q \to \overline f,t)}{N(B^0_q \to f,t)+N(\overline B^0_q \to \overline f,t)}=\\
\nonumber A_D + \frac{A_{sl}^q}{2}-\left(A_P + \frac{A_{sl}^q}{2}\right)\frac{\cos(\Delta m_q t)}{\cosh(\Delta \Gamma_q t/2)}
\end{eqnarray}
where $A_P$ and $A_D$ are the $B$-meson production asymmetry and the combined decay products detection asymmetry. For \Bs, the time-dependent term cancels due to the fast oscillations. In the \Bd case, a time-depend fit is necessary.

Using $\Bd\to D^-(\to K\pi\pi) \mu^+ \nu X$, $\Bd\to D^{*-}(\to D^0(K\pi)\pi) \mu^+ \nu X$, and $\Bs\to D_s^-(\to KK\pi) \mu^+ \nu X$ decay chains, the \lhcb measurements are:
\begin{center}
$A_{sl}^d =(-0.02\pm0.19(\mathrm{stat})\pm0.30(\mathrm{syst}))\%$\\
$A_{sl}^s =(-0.06\pm0.50(\mathrm{stat})\pm0.36(\mathrm{syst}))\%$
\end{center}
where the systematic (second) uncertainty is dominated by the size of the control samples used in the determination of the detection asymmetry. These numbers are compatible with the Standard Model-based expectations \cite{lenz_asl}.

The angle involved in the mixing-induced \CP violation in decay $\phi_M^s\equiv\phi_s$ can be determined using the $b\to c\overline c s$ decays $\Bs\to\jpsi\Kp\Km$ \cite{LHCb-PAPER-2014-059} and $\Bs\to\jpsi\pip\pim$ \cite{LHCb-PAPER-2013-069}, as \CP violation in decay is known to be negligible for this class of decays. In the Standard Model,  $\phi_s=2\beta_s=-2\mathrm{arg}\left(\frac{-V_{ts}V_{tb}^*}{V_{cs}V_{cb}^*}\right)=-0.036\pm0.001$ \cite{ckmfitter}.
For both cases, a description of resonant and non-resonant contributions is performed and an angular analysis is needed to unfold the different \CP states.

For $\Bs\to\jpsi\pip\pim$, a fit using values of $\Gamma_s$, $\Delta\Gamma_s$ and $\Delta m_s$ constrained to known accurate experimental measurements leads to $\phi_s = 0.070\pm0.068(\mathrm{stat})\pm0.008(\mathrm{syst})$. In the case of $\Bs\to\jpsi\Kp\Km$, a global fit floating all the mixing and \CP violation parameters has been performed, and allowed to derive in particular the value $\phi_s = -0.058\pm0.049(\mathrm{stat})\pm0.006(\mathrm{syst})$. The combination yields $\phi_s = -0.010\pm0.039$, where the uncertainty is all inclusive.~This is the most precise measurement up to date, agreeing with the expectations at this level of accuracy.
\section{\CP-violating patterns in charmless three-body decays}
The three-body decays, the mesonic $\Bpm\to h^{\prime+}h^{\prime-}h^\pm$ and baryonic $\Bpm\to p \overline p h^\pm$ modes ($h,h^{\prime}=\pi,K$), share similar quark-level diagrams as the charmless two-body $B\to hh^{\prime}$ decays used for the extraction of the angle $\gamma$ (section \ref{sec:gamma_angle}). Their interesting feature lies in the possibility to study the dynamics of the amplitude, as the density of events in the Dalitz plane \cite{Dalitz} is directly proportional to the squared modulus of the amplitude $A$.~Near-threshold distributions are observed for both mesonic and baryonic cases.~However, while in the mesonic modes \cite{LHCb-PAPER-2014-044}, resonances such as $\rho$, $f_0(980)$ or \Kstar largely contribute to the spectrum, the baryonic decays \cite{LHCb-PAPER-2014-034} exhibit a non-resonant near-$\proton\antiproton$ threshold enhancement. In addition, large angular asymmetries are observed in the distribution of the helicity angle of the light meson $h$, with opposite sign between $\Bpm\to \proton \antiproton\Kpm$ and $\Bpm\to \proton \antiproton\pipm$.

\begin{figure}[t]
\begin{center}
\includegraphics[width=0.4\textwidth]{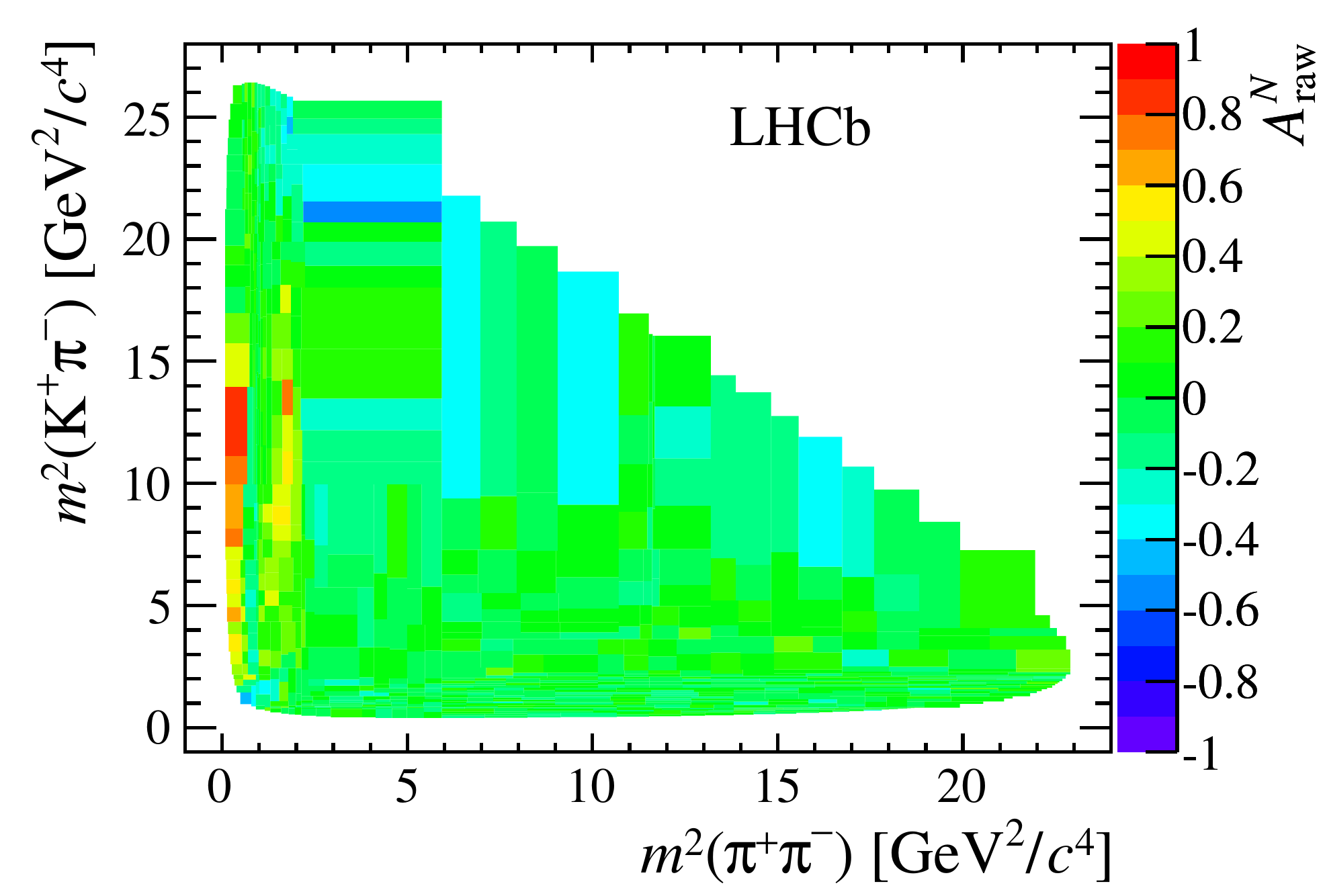}
\includegraphics[width=0.4\textwidth]{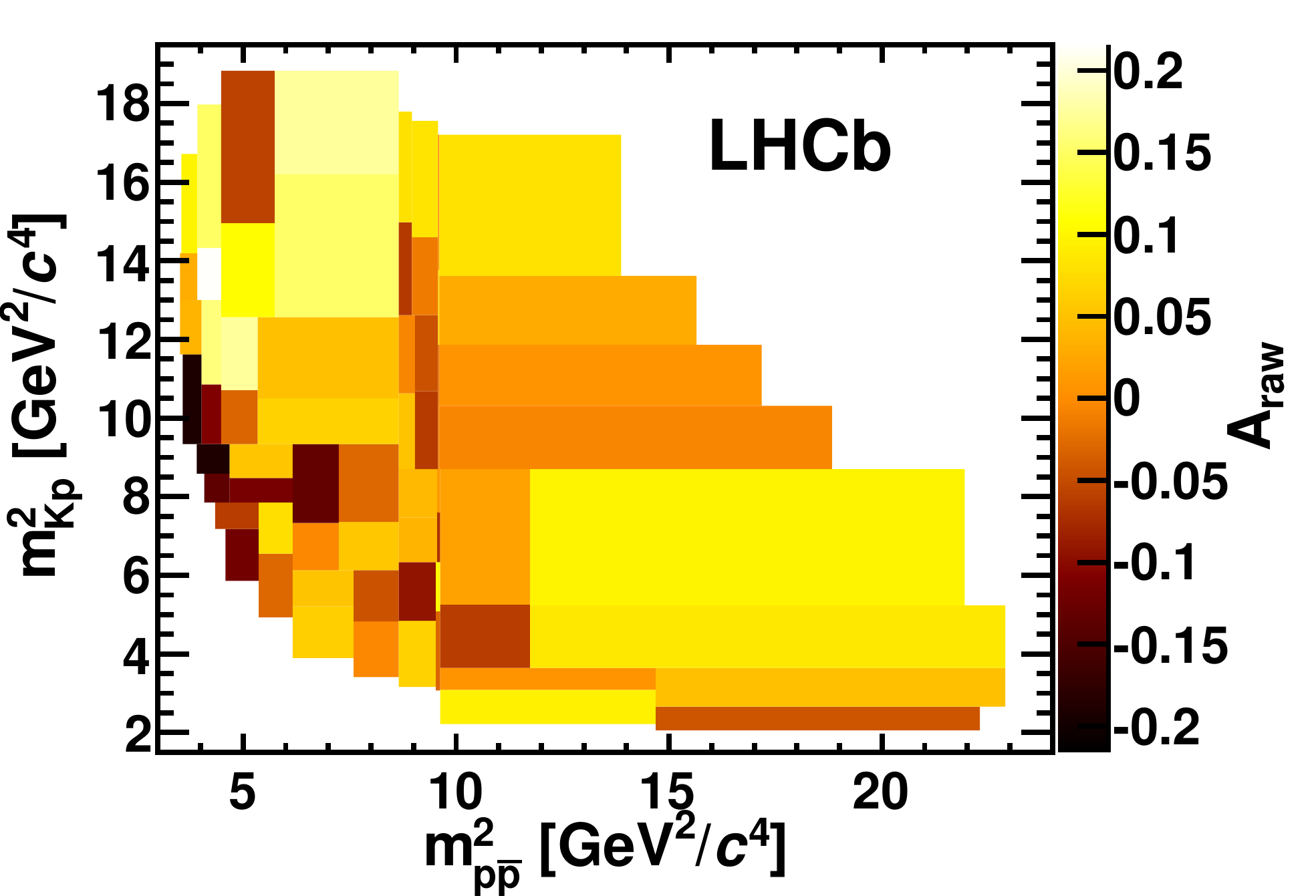}
\end{center}
\caption{Raw charge asymmetry distributions throughout the Dalitz plane for $\Bpm\to \Kpm \pi^+\pi^-$ (top) and $\Bpm\to \proton\antiproton\Kpm$ (bottom) decays. A dynamical binning is used so that the number of events contained in each bin is constant.}
\label{Fig:araw_dal_hhh}
\end{figure}

For the charge asymmetry analysis, it is known that at least two contributing amplitudes $A_1$ and $A_2$, with different strong $\delta_i$ and weak $\varphi_i$ phases, are necessary to generate a charge difference:
\begin{eqnarray}
|A(\Bm\to f^-)|^2 - |\overline A(\Bp\to f^+)|^2 =\\
\nonumber -4|A_1||A_2|\sin(\delta_1-\delta_2)\sin(\varphi_1-\varphi_2).
\end{eqnarray}
Dalitz scans of the charge asymmetry have shown peculiar \CP violating structure at low $m(h^{\prime+}h^{\prime-})$ and $m(p\overline p)$ masses, as illustrated in Fig.\ref{Fig:araw_dal_hhh}, with large asymmetries in particular for the mesonic modes. In this latter case, the observed patterns above the masses of $\rho$ or $f_0(980)$ resonances indicate possible exchanges between $\Bpm\to \pip\pim h^\pm$ and $\Bpm\to \Kp\Km h^\pm$ decays through Final State Interactions involving $\pip\pim\leftrightarrow\Kp\Km$ scattering.
For the baryonic modes, the \CP asymmetry has reached the 4$\sigma$ level for the first time for final states containing baryons, and the observed structure points at complex interferences of non-resonant $p\overline p$ waves.

\section{Search for New Physics in rare $B$-meson decays}
\begin{figure}[t]
\begin{center}
\includegraphics[width=0.35\textwidth]{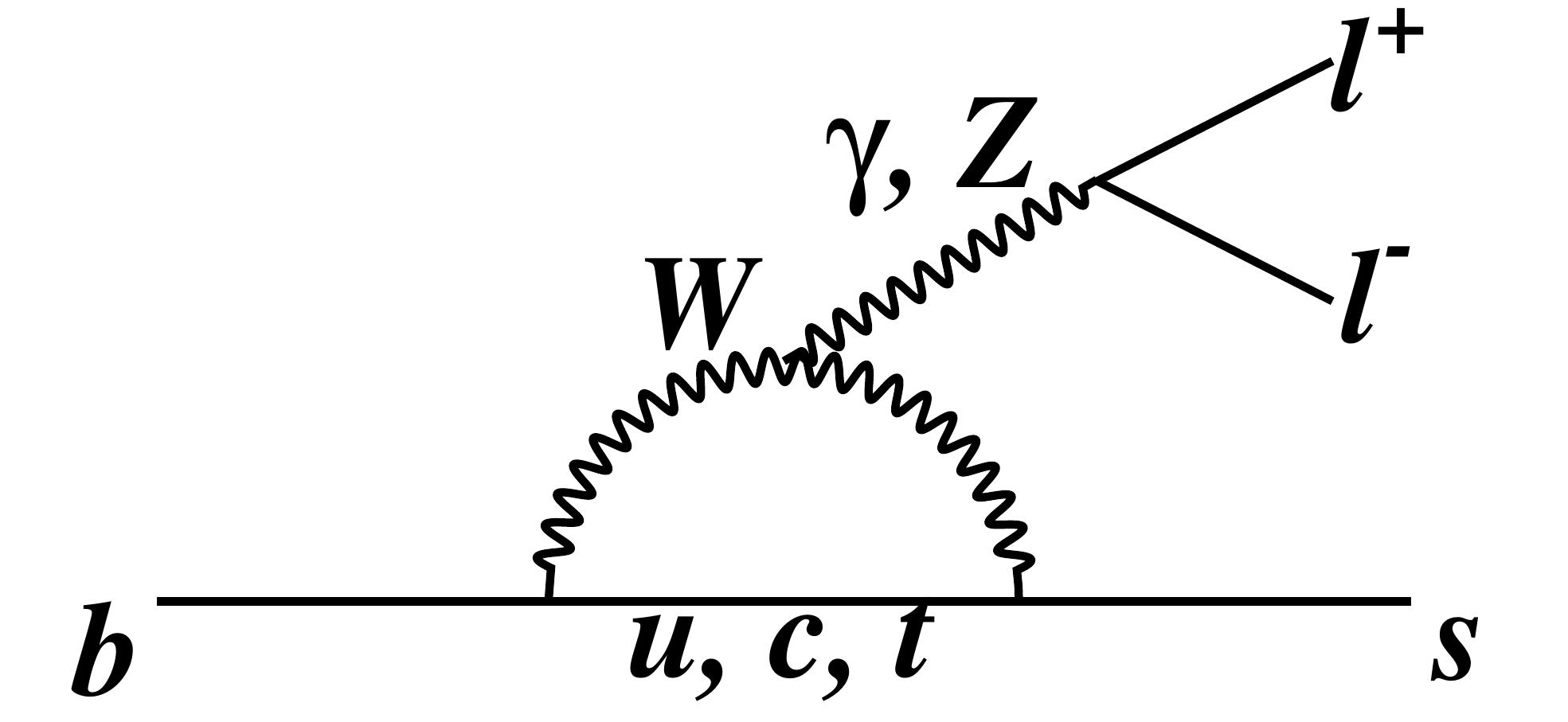}
\end{center}
\caption{Flavour changing electroweak $b\to s$ transition. The $b\overline s$ annihilation diagram is inferred by moving the s quark branch to the left.}
\label{Fig:btosll}
\end{figure}

The rare processes $B\to \ell^+\ell^-$ and $B\to K^{(*)}\ell^+\ell^-$ are mediated by electroweak loop transitions such as the one shown in Fig.\ref{Fig:btosll}.~They are therefore largely suppressed in the Standard Model.~Intervention of new heavy particles in the loop could change the overall branching fraction or the dynamics of these decays. 
\lhcb has investigated $B_{s,d}^0\to\mu^+\mu^-$ \cite{LHCb-PAPER-2013-046}, \BdToKstmm \cite{LHCb-PAPER-2013-037} and $B^\pm\to\Kpm\ell^+\ell^-$ \cite{LHCb-PAPER-2014-024}.

For $B_{s,d}^0\to\mu^+\mu^-$, a combination of the results with \cms collaboration \cite{LHCb-PAPER-2014-049} has permitted the first observation of $B_{s}^0\to\mu^+\mu^-$, $\mathcal{B}(B_{s}^0\to\mu^+\mu^-)=(2.8^{+0.7}_{-0.6})\times 10^{-9}$, and an evidence for $B_{d}^0\to\mu^+\mu^-$, with significances of 6.2$\sigma$ and 3.2$\sigma$, respectively.~The ratio $\mathcal{B}(B_{d}^0\to\mu^+\mu^-)/\mathcal{B}(B_{s}^0\to\mu^+\mu^-)$ exceeds the Standard Model prediction by 2.3$\sigma$.

In the case of $B^\pm\to\Kpm\ell^+\ell^-$, lepton universality requires equal branching fractions for $B^\pm\to\Kpm e^+e^-$ and $B^\pm\to\Kpm \mu^+\mu^-$. The analyses have performed an integration of the rates in the region free from charmonium events, $1<q^2\equiv m^2(\ell^+\ell^-)<6\gevgevcccc$, and found $R_K=\frac{\int_{q^2_{min}}^{q^2_{max}} d\Gamma(B^\pm\to\Kpm \mu^+\mu^-)/dq^2}{\int_{q^2_{min}}^{q^2_{max}} d\Gamma(B^\pm\to\Kpm e^+e^-)/dq^2}$ to be $0.745^{+0.090}_{-0.074}\pm0.036$, which is 2.6$\sigma$ from unity.

The study of $\Bd\to\Kstarz(\Km\pip)\mup\mun$ includes an angular analysis involving the direction $\theta_l$ of the $\mu\mu$ decay with respect to the $B$ meson, the polarization angle $\theta_K$ of \Kstarz and the angle $\phi$ between the $\mu\mu$ and $K\pi$ decay planes. Nor the integrated branching fraction neither the variation of the $\mu\mu$ Forward-Backward asymmetry as a function of $q^2=m^2(\mu^+\mu^-)$ have shown any sign of noticeable discrepancy compared to the expectations. However, a puzzling 3.7$\sigma$ deviation observed for one coefficient of the angular distribution in one bin of the $q^2$ variable is still subject to investigations and will be further clarified in a coming updated version of the analysis.
\section{Heavy hadrons}
\lhcb is also very active in studying heavy hadrons spectroscopy and has various ongoing analyses in this area of research.
\subsection{\Bc meson decays}
\begin{figure}[t]
\begin{center}
\includegraphics[width=0.4\textwidth]{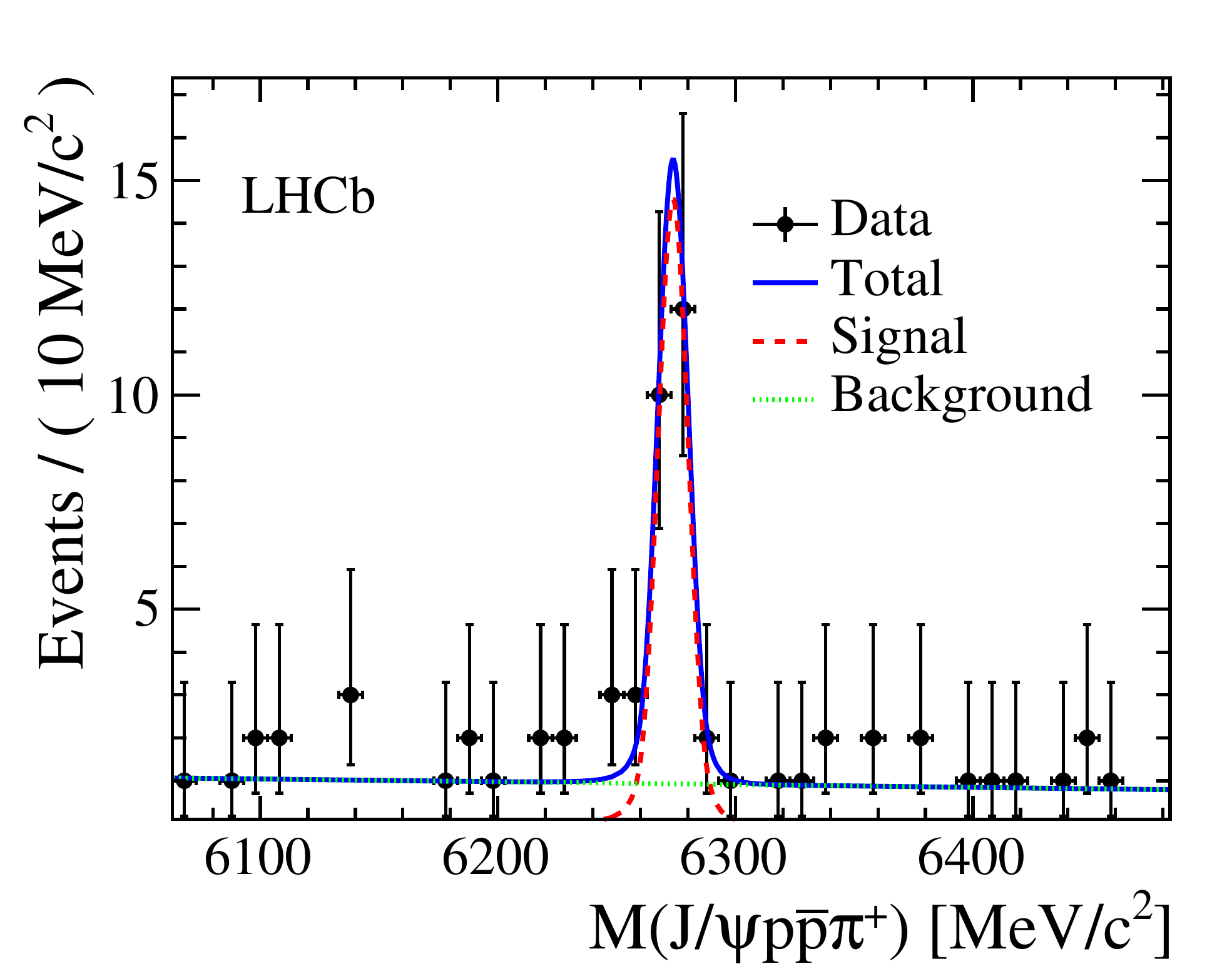}
\includegraphics[width=0.4\textwidth]{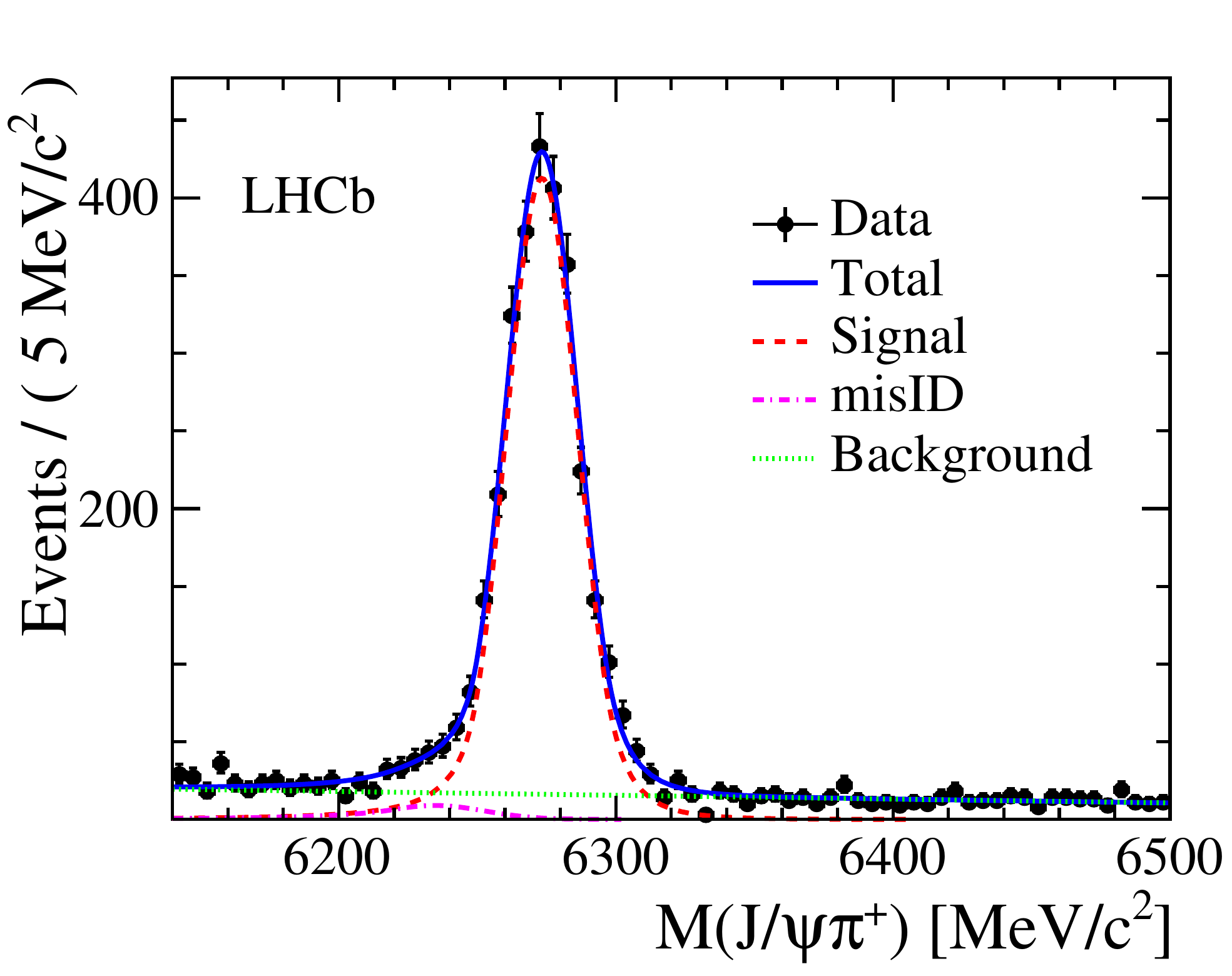}
\end{center}
\caption{Invariant mass distribution of $\jpsi\proton\antiproton\pip$ (top) and $\jpsi\pip$ (bottom) combinations, with the clear \Bc signal peaks.}
\label{Fig:bctojpsiX}
\end{figure}

The \Bc meson ($\overline bc$) has the specificity of being a doubly heavy flavoured meson and thus has the two important properties of having a mass $\sim$ 1 \gevcc higher than all the other $B$ mesons and decaying through either $b$ or $c$ quark decays. Due to the dominant CKM matrix element $V_{cs}\sim 1$, about 70\% of the \Bc decays proceed through \cquark decays, leading to intermediate states of the type $B_q\pi$ or $B_qK$ ($q=d,s$). The remainder of the decay width is dominated by $b\to c$ transitions leading to final states with charmonium ($X_{c\overline c}$), either semileptonic $\Bc\to X_{c\overline c} \ell^+\nu$ or non-leptonic $\Bc\to X_{c\overline c} X$. Finally, charmless $b$ quark decays to $u,s,d$, leading to final states with one charm $D$ meson, are suppressed, and the $\overline bc$ annihilation process, generating final states with no charmed hadron, is even more suppressed. In \lhcb, several analyses have explored \Bc properties such as the lifetime \cite{LHCb-PAPER-2013-063,LHCb-PAPER-2014-060}, or its production in $pp$ collisions \cite{LHCb-PAPER-2014-050}. Among the recent results, special attention may be drawn to the first observation of a \Bc decay with a final state containing baryons, $\Bc\to\jpsi\proton\antiproton\pip$ \cite{LHCb-PAPER-2014-039}. The corresponding mass distribution, along with the one for the control channel $\Bc\to\jpsi\pip$, is shown in Fig.\ref{Fig:bctojpsiX}. After correcting for detection and selection efficiencies, the following relative branching fraction is obtained:
\begin{center}
$\frac{\mathcal{B}(\Bc\to\jpsi\proton\antiproton\pip)}{\mathcal{B}(\Bc\to\jpsi\pip)}=0.143^{+0.039}_{-0.034}\mathrm{(stat)}\pm0.013\mathrm{(syst)}$
\end{center}

\subsection{Observation of two new $\Xi_b$ states}
\begin{figure}[t]
\begin{center}
\includegraphics[width=0.35\textwidth]{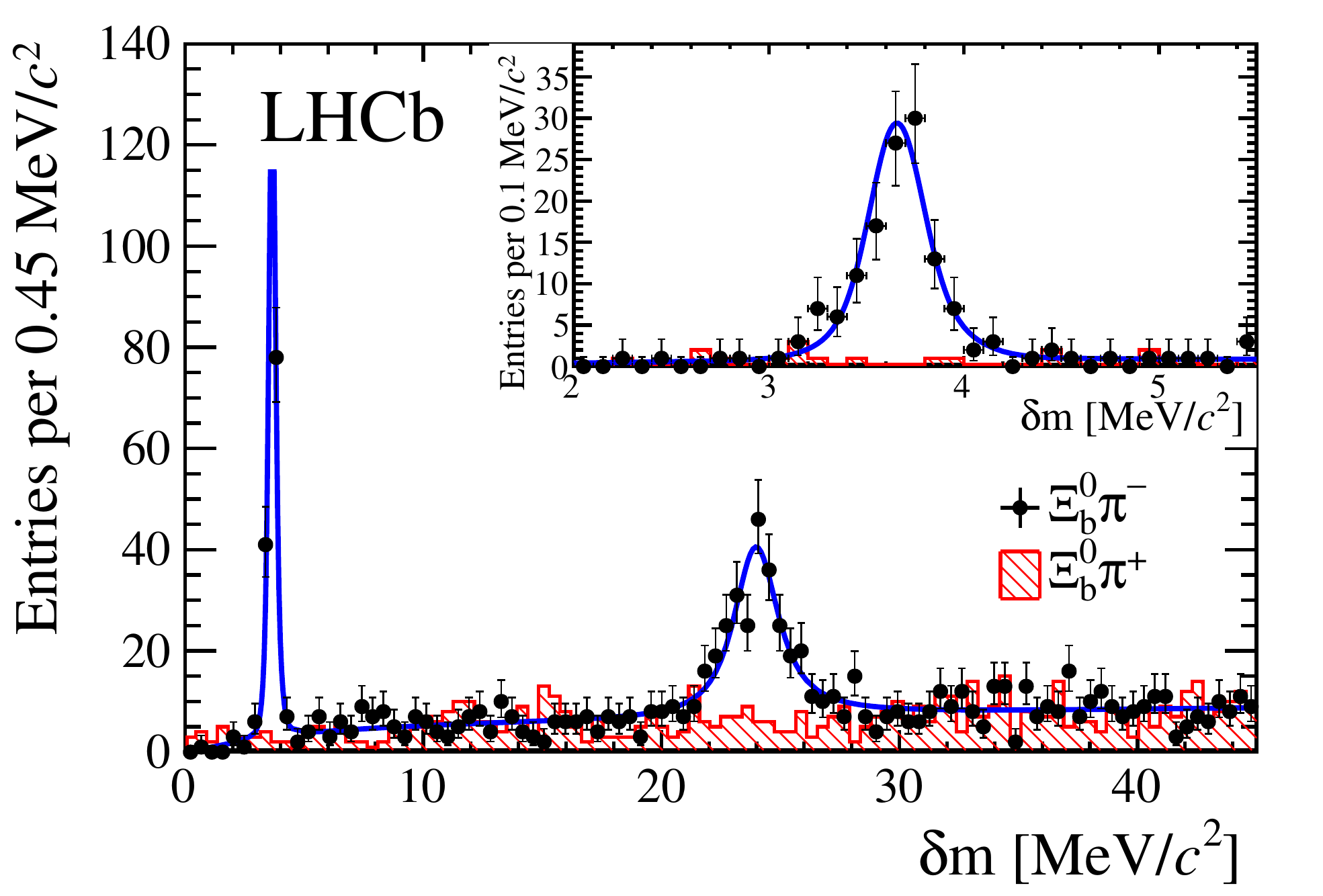}
\end{center}
\caption{Distribution of $\delta m = m(\Xi_b^0\pim)-m(\Xi_b^0)-m_\pim$. $\Xi_b^0$ is reconstructed in the mode $\Xi_b^0\to\Xi_c^+(\to p \Km\pip)\pim$. The hatched histogram represents the wrong-sign combinations. The inset shows the zoom in the region $2.0-5.5\mevcc$}
\label{Fig:xsib_signals}
\end{figure}

 Recently \cite{LHCb-PAPER-2014-061}, two excited states of the strange $\Xi_b^+$ baryon ($\overline b\overline s \overline d$) have been observed in an analysis where $\Xi_b^0\to\Xi_c^+(\to p \Km\pip)\pim$ candidates are combined with charged pions. The $\Xi_b^0\pi$ spectrum, Fig.\ref{Fig:xsib_signals}, shows two clear peaks for which the investigations have proven to be resonances of masses $m(\Xi_b^\prime)=5935.02\pm0.02\pm0.01\pm0.50\mevc$ and $m(\Xi_b^*)=5955.33\pm0.12\pm0.06\pm0.50\mevc$, and widths $\Gamma(\Xi_b^\prime)<0.08\mevc$ at 95\% confidence level and $\Gamma(\Xi_b^*)=1.65\pm0.31\pm0.10\mevc$, respectively.~The first and second uncertainties are statistical and systematic, while the third one (for the masses) is related to the precision on the mass of $\Xi_b^0$.

\section{Forward physics studies}
The complementarity in pseudo-rapidity coverage of the \lhcb detector with the \atlas and \cms detectors has triggered several non-heavy-flavour physics analyses including topics on QCD, electroweak physics, Higgs boson and search for new heavy long-lived particles. Beside first searches such as the search for the decay $\H\to\tau^+\tau^-$ \cite{LHCb-PAPER-2013-009}, more recent studies include the measurement of the $Z+b$-jet production cross-section and a search for a hypothetical heavy long-lived particle decaying into a $b\overline b$ pair. For $Z+b$-jet \cite{LHCb-PAPER-2014-055}, cross sections in the range $100-300\fb$ have been measured for two $b$-jet transverse momentum thresholds, $\pt>10,~20\gevc$, with substantial statistical and systematical uncertainties. 

\begin{figure}[t]
\begin{center}
\includegraphics[width=0.4\textwidth]{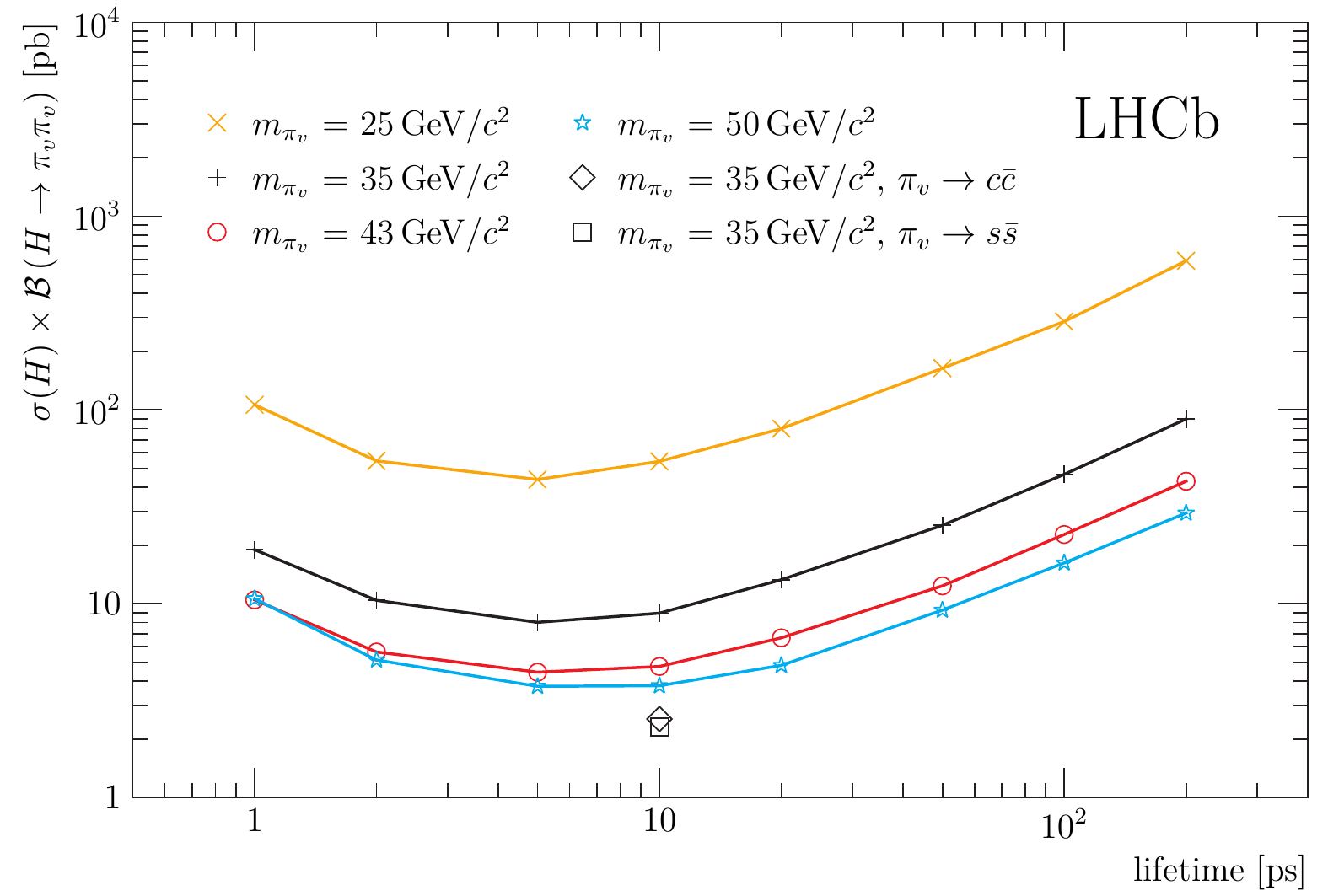}
\end{center}
\caption{95\% confidence level upper limits on $\pi_v$ production cross-section from Higgs boson decays, for various $\pi_v$ masses, as a function of lifetime. The $\pi_v\to b\bar{b}$ decays are considered by default, but examples for $\pi_v\to c\bar{c},~s\bar{s}$ are also quoted.}
\label{Fig:vpion_xsec}
\end{figure}

 Within Hidden Valley models \cite{HV1}, some configurations predict the existence of a hypothetical $v$-pion, $\pi_v$, produced in neutral Higgs boson decays $H\to\pi_v\pi_v$, and decaying into $b \overline b$ pairs (and other $q\overline q$ decays). A \lhcb study \cite{LHCb-PAPER-2014-062} has been able to set upper limits on the production cross-section of $\pi_v$, as a function of its mass and lifetime, as can be seen in Fig. \ref{Fig:vpion_xsec}.
\section{Conclusions}
The analysis of Run I data has permitted various improvements in the measurements of \CP-violation parameters in $B$ mesons oscillations and decays: the precision on the CKM angles $\gamma$ and $\beta_s$ from the sole \lhcb studies is now below $10^\circ$ and 2\%, respectively, and is expected to reach less than half these values with Run II data. A huge number of observations of new particles, excited states and decays has been published. The observation of the rare $\Bs\to\mup\mun$ decay is also one of the main achievements. Hints of discrepancies in observables of the rare $B$ decays have been remarked but are still lacking significance. \CP violation studies in multibody decays have unveiled complex final state dynamics that still needs to be understood in amplitude analyses. Forward physics, including the search of Higgs-boson and long-lived heavy particles decays, has started to provide first results and might be a corner from where surprises emerge in the future.





\addcontentsline{toc}{section}{References}
\setboolean{inbibliography}{true}
\bibliographystyle{LHCb}
\bibliography{mybib,LHCb-PAPER,LHCb-CONF,LHCb-DP}

\end{document}

%% file: lhcb-symbols-def.tex
\def\lhcb {\mbox{LHCb}\xspace}
\def\atlas  {\mbox{ATLAS}\xspace}
\def\cms    {\mbox{CMS}\xspace}

\ifthenelse{\boolean{uprightparticles}}%
{

 \def\Pmu         {\ensuremath{\upmu}\xspace}

 \def\Ppi         {\ensuremath{\uppi}\xspace}

 \def\Ppsi        {\ensuremath{\uppsi}\xspace}

 \def\PDelta      {\ensuremath{\Delta}\xspace}                 
 \def\PXi      {\ensuremath{\Xi}\xspace}                 
 \def\PLambda      {\ensuremath{\Lambda}\xspace}                 
 \def\PSigma      {\ensuremath{\Sigma}\xspace}                 
 \def\POmega      {\ensuremath{\Omega}\xspace}                 
 \def\PUpsilon      {\ensuremath{\Upsilon}\xspace}                 
 
 \def\PB      {\ensuremath{\mathrm{B}}\xspace}                 
                  
 \def\PD      {\ensuremath{\mathrm{D}}\xspace}

 \def\PH      {\ensuremath{\mathrm{H}}\xspace}                 
                  
 \def\PJ      {\ensuremath{\mathrm{J}}\xspace}                 
 \def\PK      {\ensuremath{\mathrm{K}}\xspace}

 \def\Pc      {\ensuremath{\mathrm{c}}\xspace}

 \def\Pi      {\ensuremath{\mathrm{i}}\xspace}

 \def\Pp      {\ensuremath{\mathrm{p}}\xspace}

 \def\Ps      {\ensuremath{\mathrm{s}}\xspace}

}
{

 \def\Pmu         {\ensuremath{\mu}\xspace}

 \def\Ppi         {\ensuremath{\pi}\xspace}

 \def\Ppsi        {\ensuremath{\psi}\xspace}                 
                  
 \mathchardef\PDelta="7101
 \mathchardef\PXi="7104
 \mathchardef\PLambda="7103
 \mathchardef\PSigma="7106
 \mathchardef\POmega="710A
 \mathchardef\PUpsilon="7107
                  
 \def\PB      {\ensuremath{B}\xspace}                 
                  
 \def\PD      {\ensuremath{D}\xspace}

 \def\PH      {\ensuremath{H}\xspace}                 
                  
 \def\PJ      {\ensuremath{J}\xspace}                 
 \def\PK      {\ensuremath{K}\xspace}

 \def\Pc      {\ensuremath{c}\xspace}

 \def\Pi      {\ensuremath{i}\xspace}

 \def\Pp      {\ensuremath{p}\xspace}

 \def\Ps      {\ensuremath{s}\xspace}

}

\makeatletter
\ifcase \@ptsize \relax
  \newcommand{\miniscule}{\@setfontsize\miniscule{4}{5}}
\or
  \newcommand{\miniscule}{\@setfontsize\miniscule{5}{6}}
\or
  \newcommand{\miniscule}{\@setfontsize\miniscule{5}{6}}
\fi
\makeatother

\DeclareRobustCommand{\optbar}[1]{\shortstack{{\miniscule (\rule[.5ex]{1.25em}{.18mm})}
  \\ [-.7ex] $#1$}}

\def\mup        {{\ensuremath{\Pmu^+}}\xspace}
\def\mun        {{\ensuremath{\Pmu^-}}\xspace} 

\def\H      {{\ensuremath{\PH^0}}\xspace}

\def\squark    {{\ensuremath{\Ps}}\xspace}

\def\cquark    {{\ensuremath{\Pc}}\xspace}

\def\pion   {{\ensuremath{\Ppi}}\xspace}

\def\pip    {{\ensuremath{\pion^+}}\xspace}
\def\pim    {{\ensuremath{\pion^-}}\xspace}
\def\pipm   {{\ensuremath{\pion^\pm}}\xspace}

\def\kaon    {{\ensuremath{\PK}}\xspace}
  \def\Kbar    {{\kern 0.2em\overline{\kern -0.2em \PK}{}}\xspace}

\def\KorKbar    {\kern 0.18em\optbar{\kern -0.18em K}{}\xspace}

\def\Kp      {{\ensuremath{\kaon^+}}\xspace}
\def\Km      {{\ensuremath{\kaon^-}}\xspace}
\def\Kpm     {{\ensuremath{\kaon^\pm}}\xspace}

\def\KS      {{\ensuremath{\kaon^0_{\rm\scriptscriptstyle S}}}\xspace}

\def\Kstarz  {{\ensuremath{\kaon^{*0}}}\xspace}

\def\Kstar   {{\ensuremath{\kaon^*}}\xspace}

  \def\Dbar    {{\kern 0.2em\overline{\kern -0.2em \PD}{}}\xspace}

\def\DorDbar    {\kern 0.18em\optbar{\kern -0.18em D}{}\xspace}

\def\B       {{\ensuremath{\PB}}\xspace}
\def\Bbar    {{\ensuremath{\kern 0.18em\overline{\kern -0.18em \PB}{}}}\xspace}

\def\BorBbar    {\kern 0.18em\optbar{\kern -0.18em B}{}\xspace}

\def\Bu      {{\ensuremath{\B^+}}\xspace}
\def\Bub     {{\ensuremath{\B^-}}\xspace}
\def\Bp      {{\ensuremath{\Bu}}\xspace}
\def\Bm      {{\ensuremath{\Bub}}\xspace}
\def\Bpm     {{\ensuremath{\B^\pm}}\xspace}

\def\Bd      {{\ensuremath{\B^0}}\xspace}
\def\Bs      {{\ensuremath{\B^0_\squark}}\xspace}

\def\Bc      {{\ensuremath{\B_\cquark^+}}\xspace}

\def\jpsi     {{\ensuremath{{\PJ\mskip -3mu/\mskip -2mu\Ppsi\mskip 2mu}}}\xspace}

  \def\Y#1S{\ensuremath{\PUpsilon{(#1S)}}\xspace}

\def\proton      {{\ensuremath{\Pp}}\xspace}
\def\antiproton  {{\ensuremath{\overline \proton}}\xspace}

\def\Lbar        {{\ensuremath{\kern 0.1em\overline{\kern -0.1em\PLambda}}}\xspace}
\def\LorLbar    {\kern 0.18em\optbar{\kern -0.18em \PLambda}{}\xspace}

\newcommand{\decay}[2]{\ensuremath{#1\!\to #2}\xspace}         

\def\to                 {\ensuremath{\rightarrow}\xspace}

\def\CP                {{\ensuremath{C\!P}}\xspace}

\def\BdToKstmm    {\decay{\Bd}{\Kstarz\mup\mun}}

\def\AT#1     {\ensuremath{A_{\mathrm{T}}^{#1}}\xspace}           

\def\C#1      {\ensuremath{\mathcal{C}_{#1}}\xspace}                       
\def\Cp#1     {\ensuremath{\mathcal{C}_{#1}^{'}}\xspace}                    
\def\Ceff#1   {\ensuremath{\mathcal{C}_{#1}^{\mathrm{(eff)}}}\xspace}        
\def\Cpeff#1  {\ensuremath{\mathcal{C}_{#1}^{'\mathrm{(eff)}}}\xspace}       
\def\Ope#1    {\ensuremath{\mathcal{O}_{#1}}\xspace}                       
\def\Opep#1   {\ensuremath{\mathcal{O}_{#1}^{'}}\xspace}                    

\newcommand{\tev}{\ifthenelse{\boolean{inbibliography}}{\ensuremath{~T\kern -0.05em eV}\xspace}{\ensuremath{\mathrm{\,Te\kern -0.1em V}}}\xspace}
\newcommand{\gev}{\ensuremath{\mathrm{\,Ge\kern -0.1em V}}\xspace}
\newcommand{\mev}{\ensuremath{\mathrm{\,Me\kern -0.1em V}}\xspace}
\newcommand{\kev}{\ensuremath{\mathrm{\,ke\kern -0.1em V}}\xspace}
\newcommand{\ev}{\ensuremath{\mathrm{\,e\kern -0.1em V}}\xspace}
\newcommand{\gevc}{\ensuremath{{\mathrm{\,Ge\kern -0.1em V\!/}c}}\xspace}
\newcommand{\mevc}{\ensuremath{{\mathrm{\,Me\kern -0.1em V\!/}c}}\xspace}
\newcommand{\gevcc}{\ensuremath{{\mathrm{\,Ge\kern -0.1em V\!/}c^2}}\xspace}
\newcommand{\gevgevcccc}{\ensuremath{{\mathrm{\,Ge\kern -0.1em V^2\!/}c^4}}\xspace}
\newcommand{\mevcc}{\ensuremath{{\mathrm{\,Me\kern -0.1em V\!/}c^2}}\xspace}

\def\mum  {\ensuremath{{\,\upmu\rm m}}\xspace}

\def\fb   {\ensuremath{\mbox{\,fb}}\xspace}
\def\invfb   {\ensuremath{\mbox{\,fb}^{-1}}\xspace}

\def\gsim{{~\raise.15em\hbox{$>$}\kern-.85em
          \lower.35em\hbox{$\sim$}~}\xspace}
\def\lsim{{~\raise.15em\hbox{$<$}\kern-.85em
          \lower.35em\hbox{$\sim$}~}\xspace}

\def\pt         {\mbox{$p_{\rm T}$}\xspace}

\def\evtgen     {\mbox{\textsc{EvtGen}}\xspace}

\def\geant      {\mbox{\textsc{Geant4}}\xspace}

\def\photos     {\mbox{\textsc{Photos}}\xspace}

\def\pythia     {\mbox{\textsc{Pythia}}\xspace}

\def\tell1  {TELL1\xspace}
\def\ukl1   {UKL1\xspace}

